\documentstyle[stwol,epsfig]{article}

\def\bee{\begin{equation}}
\def\ee{\end{equation}}
\def\bea{\begin{eqnarray}}
\def\eea{\end{eqnarray}}

\begin{document}

\title{DIRECT SOLUTION OF THE HARD POMERON PROBLEM FOR \newline 
 ARBITRARY CONFORMAL WEIGHT}

\author{J. WOSIEK and R. A. JANIK}

\address{Institute of Physics, Jagellonian University,\\
Reymonta 4, 30-059 Cracow, Poland}

\twocolumn[\maketitle\abstracts{
 A new method is applied to solve the Baxter equation for the one
dimensional system of
noncompact spins. Dynamics of such an ensemble is equivalent to that of a
set of reggeized gluons exchanged in the high energy limit of QCD amplitudes.
The technique offers more insight into the old calculation of the intercept
of the hard Pomeron and provides new results in the odderon channel.
}]
Quantitative description of the reggeization of QCD 
 still remains a challenge for the Leading Logarithmic
scheme and its extensions \cite{firsta,firstb}. 
In the first approximation the problem  separates into sectors with 
fixed number $n$ of the reggeized gluons propagating in the $t$ channel.
The lowest nontrivial case, $n=2$, was solved in the classical papers 
by Balitskii,
Kuraev, Fadin and Lipatov \cite{BFKL}
resulting in the simple expression for the intercept of the hard pomeron.
The notable progress for arbitrary $n$ was achieved by Lipatov and 
Faddeev and Korchemsky  
\cite{LIP0,FK} who have established
exact equivalence with the one dimensional chain of $n$ noncompact spins.
The success of this
approach was confirmed by 
rederiving
the Lipatov et al. result in the $n=2$ case \cite{FK,kor1}.
 However, the adopted procedure 
requires an analytic continuation from the integer values
of the relevant conformal weight $h$ (see later) because only for integer $h$
they were  able to diagonalize the two spin hamiltonian.  The $n=3$ case,
 which gives the lowest contribution to the 
odderon exchange, was studied by Lipatov,  Faddeev and Korchemsky, 
 \cite{LIP1,FK,kor1}.
Again, the spectrum of the system for integer $h$ can be found for any finite 
$h=m$. However, the general expression for arbitrary $m$ 
is not known, and
consequently the analytical continuation to $h=1/2$ is not available
\footnote{The lowest state of the $n=3$ hamiltonian is believed to occur
 at $h=1/2$.}.

 We have developed a new approach which a) works for arbitrary values
of the conformal weight $h$, 
providing explicitly above continuation,
and b) gives  the analytic solution of the $n=3$ problem for arbitrary
$h$ and $q_3$. Here we will apply the new method to the
$n=2$ case rederiving directly the BFKL result without need of the
analytical continuation. Our new results in the $n=3$ case \cite{we1} 
 will be also shortly summarized. 
 
The intercept of the Pomeron trajectory is given by 
\begin{equation}
\alpha_P(0)=1+{\alpha_s N_c \over 4\pi}\left(\epsilon_2(h)+
 \overline{\epsilon}_2(\overline{h})\right), \label{inter}
\end{equation}
where $\epsilon_2$ and $\overline{\epsilon}_2 $ are respectively the largest 
eigenvalues
of the $n=2$ reggeon hamiltonian and its antiholomorphic counterpart
 \cite{FK,kor1}. This system is equivalent to the 
misleadingly simple set of the
 two noncompact spins which for higher $n$ generalizes to the one
dimensional chain  with nearest-neighbour interactions. Applying
 Bethe ansatz one obtains in the $n=2$ case 
\begin{equation}
\epsilon_2=i \left({\dot{Q}_2(-i)\over Q_2(-i)}-{\dot{Q}_2(i)\over Q_2(i)} 
                            \right)-4, \label{holo}
\end{equation}
where $Q_2(\lambda)$ satisfies the following Baxter equation
\begin{equation}
(\lambda+i)^2 Q_2(\lambda+i)+(\lambda-i)^2 Q_2(\lambda-i)=
(2\lambda^2+q_2) Q_2(\lambda).  \label{bax}
\end{equation}
$q_2$  is the eigenvalue of the square of the total spin of the system 
$\hat{q}_2$.
It commutes with the hamiltonian and its spectrum is 
 known from the symmetry considerations
\begin{equation}
q_2=h(1-h),\;\;\;\;h={1\over 2}(1+m) -i\nu,\;\;  m\in Z, \nu\in R.
\label{spec}
\end{equation}
In order to solve the Baxter equation, (\ref{bax}), the following integral
representation is customarily used
\bee
Q_2(\lambda)=\int_{C_I}   z^{-i\lambda-1} (1-z)^{i\lambda+1} Q(z) dz.
\label{oldan}
\ee
Then, if the boundary terms do not contribute, Eq.(\ref{bax}) is 
equivalent to the simple hypergeometric equation for $Q(z)$
\bee
 \left[ {d\over dz}z(1-z){d\over dz} -q_2 
 \right] Q(z)=0, \label{diff}
\ee
with the well known solutions. However, for arbitrary value of the 
conformal weight,
$h$ the singularity structure of the hypergeometric functions together with
the nontrivial monodromy of the kernel 
$K(z, \lambda)=z^{-i\lambda-1}(1-z)^{i\lambda+1}$ precludes existence
of the contour such that the boundary contributions cancel. For integer
$h=m$, however, the solution regular at $z=0$ does not have a cut and
consequently the simple contour encircling both $z=0$ and $z=1$ points
guarantees vanishing of the boundary terms. This observation was exploited
in Refs{\cite{FK,kor1}} leading to the elegant solution of the $n=2$ problem
for integer conformal weight. The BFKL formula resulted after the analytic
continuation in $h$ to $h=1/2$. However, the case of noninteger $h$ requires 
further insight. In particular the boundary conditions for
$Q_2(\lambda)$ are not fully understood. For integer $h$, again, they can be
deduced from the polynomial Bethe ansatz and are consistent with the above
choice of the integration contour in Eq.(\ref{oldan}). For arbitrary $h$,
they are not available. It would be very instructive to investigate the
so called functional Bethe ansatz in this connection. 

  We will present here a different approach. 
It was observed in Ref.\cite{jan1} that the {\em double contour} 
representation (c.f. Fig.1)
\bea   
Q_2(\lambda)&=&\int_{C_I}   z^{-i\lambda-1} (1-z)^{i\lambda+1} Q_I(z) dz 
 \label{dcon} \\
  &+&\int_{C_{II}}   z^{-i\lambda-1} (1-z)^{i\lambda+1} Q_{II}(z) dz,
\nonumber
\eea
together with simple boundary conditions on $Q_{I/II}(z)$, 
reproduced numerically
the holomorphic energy in the half-integer case $h=m+1/2$. Using  the double
contour
representation we have subsequently derived the analytic expression
for the holomorphic energy  for arbitrary complex $h$. With the aid of the new
formalism of the transition matrix this method was applied to the $n=3$ case
and led to the analytic expression for the intercept of the odderon 
trajectory for arbitrary values of relevant parameters. 

We begin with 
 the general solutions of Eq.(\ref{diff}) and then show how the
original freedom is restricted leading to the unique solution.  
\begin{figure}[htb]
\vspace{9pt}
\framebox[65mm]{
\epsfxsize=6cm \epsfbox{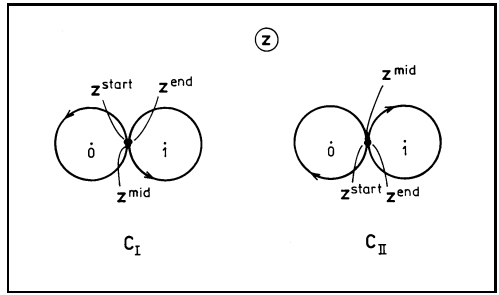}         }
\caption{ Integration contours used in Eq.(\protect\ref{dcon}). 
Start $z^{start}$,
middle $z^{mid}$, and end $z^{end}$ points coincide but they lie on the
different sheets of the Riemann surface of the integrands.
}
\label{fig:f2}
\end{figure}
To this end we write the two fundamental sets of two,
linearly independent  solutions of Eq.(\ref{diff})
\begin{eqnarray}
\vec{u}(z) = (u_1(z),u_2(z)), \\ \nonumber 
\vec{v}(z) = (v_1(z),v_2(z)), \nonumber
\end{eqnarray}
around $z=0$ and $z=1$ respectively.   
\begin{eqnarray}
u_1(z)&=& F(h,1-h,1;z)=\sum_{n=0}^{\infty}f_n z^z, \nonumber \\
u_2(z)&=& {s(h)\over \pi i}\log{z}\; u_1(z) - 
{s(h) \over \pi i}\sum_{n=0}^{\infty} g_n z^n, \label{uba} \\ 
g_n&=& 
f_n[2\psi(n+1)-\psi(n+h)-\psi(n+1-h)] ,
\nonumber
\end{eqnarray} 
where $F(a,b,c;z)$ is the hypergeometric function, $\psi(z)$ denotes
the digamma function and $s(h)=\sin{(\pi h)}$.
The series in Eq.(\ref{uba}) are convergent in the unit circle $K_0$ around
$z=0$. Similarly one can construct the $\vec{v}(z)$ solutions in the unit
circle $K_1$ around $z=1$. In fact, because of the 
symmetry of  Eq.(\ref{diff}),
 we take
\begin{equation}
v_1(z)=i u_1(1-z),\;\;\; v_2(z)=-i u_2(1-z). \label{vba}
\end{equation}
Since any solution is a
linear combination of the fundamental solutions, we have in general
\begin{eqnarray}
Q_I(z)&=&a u_1(z)+b u_2(z) \nonumber \\ 
 &\equiv & A\cdot\vec{u}(z)=A\cdot\Omega \vec{v}(z),\nonumber \\
Q_{II}(z)&=&c u_1(z)+d u_2(z) \label{abf} \\
 & \equiv &  B\cdot\vec{u}(z)=B\cdot \Omega \vec{v}(z), 
      \nonumber
\end{eqnarray}
with an obvious vector notation.   
The transition matrix $\Omega$ is defined by
\begin{equation}
\vec{u}(z)=\Omega \vec{v}(z), \label{trans}
\end{equation}
and provides the analytic continuation of our solutions $Q(z)$ between
$K_0$ and $K_1$. It plays an important role for higher $n$ and its
direct calculation for $n>2$ is rather nontrivial. 
 For the hypergeometric
equation, and for the special choice of both bases, Eqs.(\ref{uba},\ref{vba}),
$\Omega$
is very simple. Due to the identity $u_2(z)=i u_1(1-z)$ 

\begin{equation}
  \Omega=\left( \begin{array}{cc}
                       0&1 \\
                       1&0 
                \end{array}   \right). \label{omega}
\end{equation}
\begin{figure}[htb]
\vspace{9pt}
\framebox[75mm]{
\epsfxsize=4cm \epsfbox{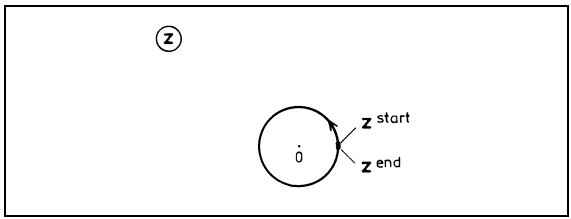} }
\caption{ Closed contour used to define the monodromy matrix, 
Eq.(\protect\ref{mon}). $z^{start}
=z^{end}$, however they belong to the different sheets of the Riemann surface.
}
\label{fig:f1}
\end{figure}
Next we introduce the monodromy matrix $M_u$
which describes the behaviour of the basis $\vec{u}$ 
in the vicinity of
 the branch point $z=0$ (see Fig.2).
\begin{equation}
\vec{u}(z_{end})=M_u \vec{u} (z_{start}), \;\;
M_u=\left( \begin{array}{cc}
                 1&0 \\
                 2s(h)&1
         \end{array} \right) , \label{mon} 
\end{equation}                                 
and similarly for the $v$ basis. It is easy to see that $M_v=M_u^{-1}$.

We are now ready to write the condition for the cancellation of the
boundary contributions in Eq.(\ref{dcon}). With the choice of the contours 
$C_I$
and $C_{II}$ as shown in Fig.1, the boundary contributions cancel if
\begin{equation}
  A^T M_I + B^T M_{II}= 0,
  \label{canc}
\end{equation}
where the combined monodromy matrices for the corresponding contours
read
\begin{equation}   
M_I=\Omega M_v \Omega^{-1} - M_u^{-1},  \;\;
M_{II}=\Omega M_v^{-1} \Omega^{-1} - M_u. \label{com}
\end{equation}
 In terms of the coefficients, condition (\ref{canc}) reads simply
\bee
a=c,\;\;\;\ b=d. \label{para}
\ee
Hence the original freedom of four coefficients in Eqs.(\ref{abf}) was reduced
to the two free parameters. In fact the energy of the system, Eq.(\ref{holo}),
is insensitive to the absolute normalization, hence only the ratio
\bee
\rho=a/b,   \label{ratio}
\ee
remains relevant. This variable parametrizes all possible boundary
conditions which are consistent with the cancellation of the end-point
contributions in the sum (\ref{dcon}). The role of the remaining freedom
is better seen when the explicit result for $\epsilon_2$ is derived.

To this end we substitue Eq.(\ref{abf}) with (\ref{para}) in (\ref{dcon}) 
and integrate
resulting expression term by term expanding $Q_{I/II}(z)$ in the $u$ basis on
$C_I$, and in the $v$ basis on $C_{II}$. Since the involved series are 
absolutely convergent in corresponding domains, the final result for
$\epsilon_2(h)$ is the analytic function of $h$. Consistent choice of the
 branches of the kernel $K(z,\lambda)$ and of $Q(z)$ must be made.
 After some calculations we obtain
\bea
\epsilon_2(h)=4\psi(1)-2\psi(h)-2\psi(1-h) \nonumber \\ -{i\pi\over s }
(\rho-\rho^{-1}) . \label{fin}
\eea

It is instructive to compare this result with the original hamiltonian
of the two spins \cite{kor1}

\bee
\hat{\cal{H}}_2=4\psi(1)-2\psi(-\hat{J}_{12})-2\psi(1+\hat{J}_{12}).
\label{ham}
\ee
where the eigenvalues of $\hat{J}_{12}$ are equal to $-h$ c.f. 
Eq.(\ref{spec}). It is now evident that the choice 
\bee
\rho=\pm 1,  \label{choice}
\ee
 gives the
correct spectrum of energies. We emphasize, however, that the additional
information was required to fix the remaining freedom. This is 
different in the $n=3$ case (see below). It is important to note that the
above choice is independent of $h$ which {\em a priori} is not guaranteed.

    Substituting Eq.(\ref{fin}), with (\ref{choice}), in Eq.(\ref{inter}),
and setting $h=\overline{h}=1/2$, we reproduce the BFKL formula 
\bee
\alpha_P(0)=1+{\alpha_s N \over \pi} 4 \log{2}.
\ee
This was also obtained in Ref.\cite{kor1} after analytic continuation
of their result from integer values of $h$. 
The difference between both 
approaches is best seen by comparing Eq.(\ref{fin}) with Eq.(6.31)
of Ref.\cite{kor1}.  It follows from the form of the hamiltonian, 
Eq.(\ref{ham}), that the complete holomorphic eigenenergy $\epsilon_2(h)$ 
 is singular also at positive integer $h$. This 
is true for our result, Eq.(\ref{fin}). On the other hand, as seen from
Eq.(\ref{inter}), in order to calculate 
 the physical intercept only  the {\em real} part of $\epsilon_2$ 
 is required. It is finite 
for positive integer $h$ and was correctly reproduced by the method of 
Faddeev and Korchemsky, c.f. Eq.(6.31) in Ref.\cite{kor1}. 

One of the ingredients of the calculation presented in
Ref.\cite{kor1} is the prescription how to fix an overall constant term
in the two spin Hamiltonian, Eq.(\ref{ham}). In the present formalism
the result (\ref{fin}) also has a freedom which is parametrized by $\rho$.
It would be interesting to see if the arbitrariness seen in both methods
had the same origin. 

     Our method can be extended to higher $n$. For $n=3$ we have carried
out this procedure explicitly \cite{we1}. The
 complete set of linearly independent solutions of the
corresponding third order differential equation was constructed. 
The transition matrix between the
$\vec{u}$ and $\vec{v}$ bases was also obtained. Since in this case there is no 
simple identity connecting linearly independent solutions, the $\Omega$
matrix is nontrivial. Remarkably it turns out that the condition
for cancellation of the end-point contributions in the double integral
representation determines {\em uniquely} the final solution of the 
Baxter equation. Existing arbitrariness 
in both transforms $Q_{I/II}(z)$ is irrelevant. 
Consequently we have obtained the holomorphic (and antiholomorphic)
energies as the analytic function of the two relevant parameters $h$
and $q_3$. The new variable $q_3$ is the eigenvalue of the second,
commuting with hamiltonian, 
observable $\hat{q}_3$ which is known but unfortunately was not 
diagonalized in spite of many attempts \cite{LIP2,jan2,kor3}.
  We have therefore mapped numerically the analytic structure of
$\epsilon_3(1/2, q_3)$ in the complex $q_3$ plane. Result is sketched
in Fig.3. The holomorphic energy has a series of poles at imaginary $q_3$
\footnote{Our definition of $q_3$ is the same as in Ref.\cite{kor1}}.
The intercept of the odderon trajectory is smaller than one
for almost all values of $q_3$ including all $q_3\in R$. 
However in the vicinities of the
poles it can be arbitrarily large. Therefore any further
conclusion about the numerical value of the $\alpha_O(0)$ depends
crucially on the spectrum of $q_3$.
                  
\begin{figure}[htb]
\vspace{9pt}
\framebox[75mm]{
\epsfxsize=5cm \epsfbox{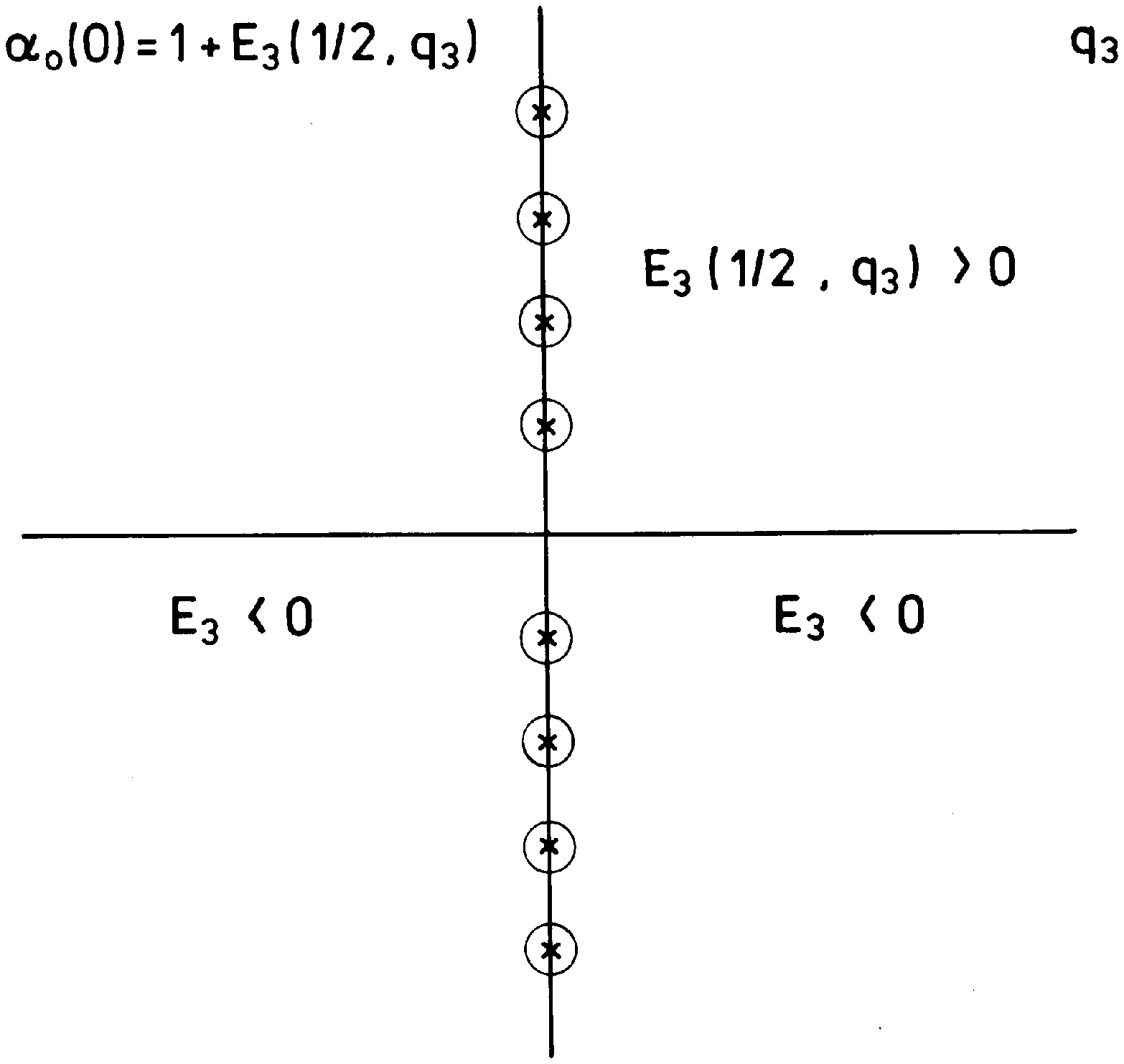} }
\caption{ Schematic map of the analyticity structure of $\protect\epsilon_3(1/2,q_3)$
in the complex $q_3$ plane. $E_3$ is positive only in the vicinity of the
poles. 
}
\label{fig:f4}
\end{figure}
We would like to thank L. N. Lipatov and G. P. Korchemsky for interesting
 discussions.
This work is supported by the Polish Committee for Scientific Research
under the grants no PB 2P03B19609 and PB 2P03B08308.    

\end{document}